\documentclass[aps,prstper,twocolumn,showpacs,letterpaper,longbibliography,nofootinbib,floatfix]{revtex4-2}

\usepackage{graphicx}
\usepackage{dcolumn}
\usepackage{bm}
\usepackage{hyperref}
\usepackage{multirow}
\usepackage{xcolor}
\usepackage{amsmath}
\usepackage{amssymb}
\usepackage[mathlines]{lineno} 

\usepackage{mathtools}

\DeclarePairedDelimiter\ket{\lvert}{\rangle}

\begin{document}

\preprint{APS/123-QED}

\title{Can a CNOT Gate Affect the Control Qubit? Student Resources for Understanding CNOT and Entanglement}

\author{Jonan-Rohi S. Plueger} \email{jopl6808@colorado.edu} \author{Bethany R. Wilcox}
\affiliation{Department of Physics, University of Colorado, Boulder, Colorado 80309, USA}
\author{Steven J. Pollock}
\affiliation{Department of Physics, University of Colorado, Boulder, Colorado 80309, USA}
\author{Gina Passante}
\affiliation{Department of Physics, California State University Fullerton, Fullerton, California 92831, USA}

\date{\today}

\begin{abstract}
The Controlled-Not (CNOT) gate is essential to algorithms in quantum computing for its ability to entangle qubits. As such, it is important to understand how students learning quantum computing reason around the function and use of this critical quantum gate. To investigate this, we conducted think-aloud interviews in which students solved problems involving the CNOT gate to understand students' `CNOT toolbox'---the strategies and cognitive resources students use when reasoning about the effect of the CNOT gate. We identify three cognitive resources related to the CNOT gate:
(1) the procedural resource of applying CNOT to specific states, (2) a qualitative description of CNOT's effect on the target qubit given the control qubit, and (3) the idea that the control qubit is not changed when CNOT is applied to computational basis states.
We find that students' use of the first resource is foundational to their understanding of the second and third, that the second and third resources can sometimes lead students to incorrect conclusions, and that students can use each of these resources separately or in tandem. We also explore how students use these resources in conjunction with Dirac notation, superposition states, and entanglement to reason both productively and unproductively about quantum computing problems. 

\end{abstract}
\maketitle

\section{Introduction}
 
The CNOT gate is an essential component of many quantum algorithms in large part due to its ability to entangle quantum bits (qubits) and is taught in almost all introductory quantum information science (QIS) classes~\cite{meyer2024introductory}. These classes, particularly those focusing primarily on the QIS subfield of quantum computing, are proliferating across academic institutions in a variety of disciplines~\cite{cervantes2021overview,plunkett2020survey,meyer2022today}, including physics, computer science~\cite{meyer2022today} and engineering~\cite{asfaw2022building,dzurak2022development}, making it a truly interdisciplinary field. In such a new but booming field, standardized curricula have yet to emerge, and physics education research (PER) has a unique opportunity to play a part in the development of quantum computing education (and, more generally, QIS education) in its nascency.

Research on student reasoning and difficulties supports curricular development that is responsive to documented observations of student behaviors. The body of work on student reasoning in quantum mechanics is large (e.g. ~\cite{marshman2015framework,emigh2020based,passante2024interactive,gire2015dirac}), but its treatment of the specific concepts relevant to quantum computing is not always relevant to the practices, norms, formalism, or interpretations of quantum computing. Recently, a smaller body of work on quantum computing-specific concepts has emerged~\cite{meyer2021states,kushimo2023dirac,bouchee2022towards}, which our study contributes to by analyzing students' reasoning about the CNOT gate. Our interest in the CNOT gate in particular is motivated both by its importance for quantum algorithms and its position as one of the few crucial two-qubit gates.

Our guiding research question in this study was: what cognitive resources do students use when reasoning about problems involving the CNOT gate? To investigate this, we conducted think-aloud interviews with students in which they were asked to respond to problems involving the CNOT gate. We analyzed their responses using a \textit{resources framework}~\cite{hammer2000student} with the goal of identifying common ideas students drew on for their solutions---what we will refer to as their `CNOT toolbox'. Our discussion of students' CNOT toolbox will also include the problem-solving methods, procedures, and heuristics we observed in our dataset, to understand the contexts in which students might activate these resources. Understanding what tools students may use or misuse when working with the CNOT gate can guide instructors in helping their students gain comfort with diverse reasoning tools, improve their tool use, and prepare to solve problems in their quantum computing careers.

We begin with an introduction to the CNOT gate (Sec.~\ref{sec:CNOT-basics}) and the literature on quantum mechanics and quantum computing reasoning relevant to our findings (Sec.~\ref{sec:quantum-research}). We then describe the context for our results, including our interview methods and questions, student demographics, our theoretical framework, and our coding process (Sec.~\ref{sec:methods}). After defining three resources (Sec.~\ref{sec:defining-resources}), one procedural and two conceptual, we analyze the content of our interviews using these resources and other problem-solving tools (Sec.~\ref{sec:discussion}). Finally, we summarize our observations on students' CNOT and quantum computing toolbox, present the limitations of our study, and suggest avenues for future researchers based on our observations (Sec.~\ref{sec:conclusion}).

\section{Background}\label{sec:background}

In this section, we first define the CNOT gate and its context in the formalism of quantum computing (Sec.~\ref{sec:CNOT-basics}), and then proceed to discuss relevant research about student reasoning in quantum physics and quantum computing (Sec.~\ref{sec:quantum-research}). Readers already familiar with quantum computing fundamentals can feel free to skip directly to Sec. \ref{sec:quantum-research}. 

\subsection{The Computational Basis and the CNOT Gate}\label{sec:CNOT-basics}

Quantum computing generalizes classical $0$ and $1$ bits to `qubits', which are quantum systems with at least two well-defined, orthogonal states such that two of the states can be easily manipulated. Qubits are most commonly written in what is known as the `computational basis', composed of the eigenstates of the Pauli Z matrix and labeled $\ket{0}$ and $\ket{1}$. Unlike a classical bit, a qubit can be in superposition states, which are normalized linear combinations of the computational basis states, e.g. $\frac{1}{\sqrt{2}}\left(\ket{0}+\ket{1}\right)$.

Systems of multiple qubits are typically expressed using the tensor product; for example, $\ket{0}\otimes\ket{0}$ represents two qubits that are both in the $\ket{0}$ state. This notation can be shortened as $\ket{00}$. In this paper, we consider the leftmost qubit the `first' qubit.
\footnote{This is not the only convention for qubit counting, and the field at large has not settled on a single convention. Another common convention is to consider the rightmost qubit the `first' qubit and the leftmost qubit the `last'.}
Multiple-qubit states can exhibit quantum entanglement, identifiable whenever a two-qubit state cannot be factored into two definite single-qubit states (this is often called `non-separability', for `separating' an $n$-qubit state into $n$ single-qubit states). For example, the unentangled state $\frac{1}{\sqrt{2}}\left(\ket{00}+\ket{10}\right)$ can be factored into $\frac{1}{\sqrt{2}}\left(\ket{0}+\ket{1}\right)\otimes\ket{0}$, but the entangled state $\frac{1}{\sqrt{2}}\left(\ket{00}+\ket{11}\right)$ cannot be factored at all.

\begin{figure}[b]
\includegraphics[width=0.2\textwidth]{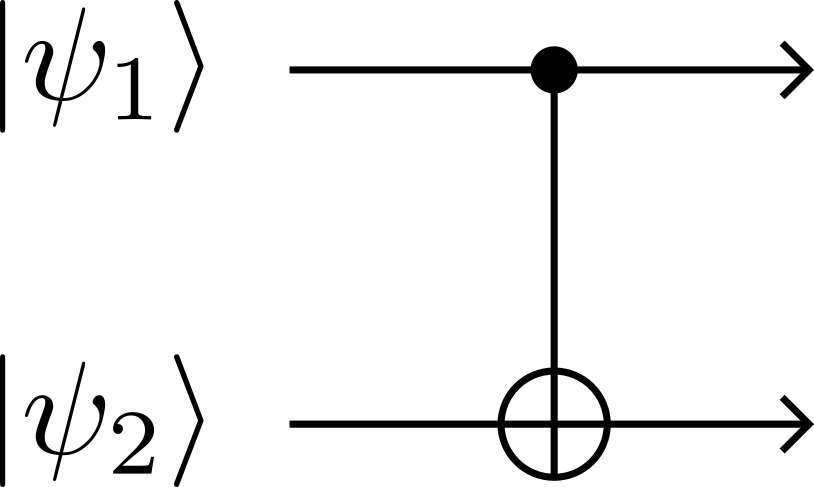}
\caption{\label{fig:CNOT-diagram}A circuit diagram for the CNOT gate, in which the top qubit is the `control qubit' and the bottom qubit is the `target qubit'. The circuit is read from left to right where the qubits are input on the left and move horizontally along the `wires'. Gates are indicated by symbols along the wires. The CNOT gate acts on two qubits and performs a NOT gate on a target qubit (in this example, the bottom qubit) based on the state of the control qubit (in this example, the top qubit).}
\end{figure}

Entangled states exhibit measurement correlation, a phenomenon in which measuring one qubit changes the probabilities that a given value will be measured for the other qubit.
The CNOT gate is critical in quantum computing for its ability to entangle two qubits by correlating their states. Figure ~\ref{fig:CNOT-diagram} provides what is known as a quantum circuit diagram representing the control qubit, target qubit, and CNOT gate for a simple two-qubit situation. In Fig.~\ref{fig:CNOT-diagram}, $\ket{\psi_1}$ is the control qubit and $\ket{\psi_2}$ is the target. The action of the CNOT in this circuit is as follows: if the control is $\ket{0}$, CNOT does nothing to the target, but if the control is $\ket{1}$, CNOT transforms the target qubit from $\ket{0} \rightarrow \ket{1}$ and $\ket{1}\rightarrow\ket{0}$ (effectively a Pauli X gate). The CNOT gate is linear and must be distributed when it acts on a superposition state such as $\frac{1}{\sqrt{2}}\left(\ket{00}+\ket{10}\right)$.
In this example, CNOT$\frac{1}{\sqrt{2}}\left(\ket{00}+\ket{10}\right) = \frac{1}{\sqrt{2}}\left(\ket{00}+\ket{11}\right)$ and the two qubits have become entangled. When entangled, the measurement of one qubit will affect the measurement probabilities of the other.

\subsection{Student Understanding of Quantum Computing and Quantum Mechanics}\label{sec:quantum-research}

Research into quantum education is a thriving field with a long history. Studies and reviews have investigated areas like student reasoning (e.g.,~\cite{bouchee2022towards,marshman2015framework,corsiglia2023intuition,krijtenburg2017insights}), curricula~\cite{emigh2020based}, online tutorials~\cite{corsiglia2022basis,corsiglia2023thesis}), simulations~\cite{passante2024interactive}, and developing assessments (e.g.,~\cite{sadaghiani2015quantum,robinett2002QMVI}). Here, we focus on the specific results of these studies that help explain student behaviors in our dataset. Some results come from studies on quantum physics education more generally, while others come from the more recent body of work on quantum computing education.

Quantum computing problems often require a fluency with `Dirac notation,' which has been the subject of previous studies. Gire and Price~\cite{gire2015dirac} observed that the use of Dirac notation resulted in fewer computational errors and enabled greater problem-solving progress than matrix and wave function notations. In that paper, they described Dirac notation as having a high level of `individuation' (clearly and distinctly representing important features of a system, such as the states in a superposition), being compact, and providing symbolic support to computations like inner and outer products. 
\footnote{Dirac notation also exposes other features of quantum states for easier manipulation or analysis, two of which are relevant to our work: (1) A state can be identified as entangled by checking if it cannot be factorized into states of individual particles (i.e., it is `non-separable' into single-qubit states). (2) The allowed results of measurement (for an observable with the same basis in which the state is expressed) can be read fairly directly from the kets themselves; especially usefully, one can read which combinations of measurements on two or more qubits are possible or not.}
However, a 2023 study on student difficulties in quantum computing~\cite{kushimo2023dirac} found that fewer than half the student participants felt comfortable with Dirac notation, preferring matrix notation instead.

\begin{table*}[]
\caption{Demographics of interview participants across multiple categories. Students were asked to answer their `gender' and `race or ethnicity... in whatever way makes sense to you'; items in those columns reflect students' exact wording. The sum of column 1 is greater than N = 29 because students reported multiple majors. }
\begin{tabular}[b]{clcc!{\vrule width 1pt}clcc!{\vrule width 1pt}clcc!{\vrule width 1pt}clcc}
\hline \hline
\; & Major \;\;\; & N & \; & \; & Academic Stage \;\;\; & N & \; & \; & Race/Ethnicity \;\;\; & N & \; & \; & Gender \;\;\; & N & \;\\
\hline
\; & Physics & 13 & \; & \; & Master's/PhD & 14 & \; & \; & Asian & 14 & \; & \; & Male & 19\\
\; & Engineering & 9 & \; & \; & Seniors & 3 & \; & \; & White & 9 & \; & \; & Female & 8\\
\; & Computer Science & 9 & \; & \; & Juniors & 5 & \; & \; & Hispanic & 3 & \; & \; & Genderqueer & 1\\
\; & Math & 8 & \; & \; & Sophomores & 4 & \; & \; & Middle Eastern & 1 & \; & \; & N/A & 1\\
\; & QIS & 5 & \; & \; & Freshmen & 3 & \; & \; & Indian & 1 & \; & \; & \multicolumn{2}{c}{}\\
\; & Economics & 1 & \; & \; & \multicolumn{3}{c!{\vrule width 1pt}}{} & \; & N/A & 1 & \; & \; \\
\hline \hline
\end{tabular}
\label{table:demographics}
\end{table*}

Some research has been done on student understanding of entanglement, especially regarding non-separability. A study by Kohnle and Deffebach~\cite{kohnle2015entanglement} on student responses based on a visualization about entanglement (in a quantum physics context) observed many students stating that a state of the form $\frac{1}{\sqrt{2}}\left(\ket{00}+\ket{01}\right)$ was `entangled because it could not be factorized,' indicating that students connect the idea of entanglement with the non-separability of a quantum state (even though, in this case, they were incorrect in saying that the state could not be factorized). However, more recently, a study by
Zwickl and Herson~\cite{zwickl2024entanglement} found that when asked to define entanglement, students rarely mentioned the idea of (non-)separability. These two results suggest that non-separability as a resource for entanglement is context-dependent.

Two recent papers~\cite{passante2015states,meyer2021states} have also uncovered student difficulties with interpreting superposition states more generally. Passante et al.\ in 2015~\cite{passante2015states} asked students whether an ensemble of particles in the superposition state $\frac{1}{\sqrt{2}}\left(\ket{0}+\ket{1}\right)$ was experimentally indistinguishable from a classical ensemble of qubits of which $50\%$ were in the $\ket{0}$ state and $50\%$ in the $\ket{1}$ state. Before direct instruction, $42\%$ of students concluded that the ensembles were experimentally indistinguishable and another $14\%$ of students claimed that the difference between the two ensembles was purely semantic.

Meyer et al.~\cite{meyer2021states} argue that the above phenomenon is a possible result of a strategy they call `Naive Measurement Probability' (NMP). In this strategy, students square the probability amplitudes on a quantum state written in the computational basis and use the resulting probabilities to make physical sense of the state. When students are asked if the quantum and classical ensembles are identical, 8 (of 15) students use NMP to explain that measurement of $\frac{1}{\sqrt{2}}\left(\ket{0}+\ket{1}\right)$ will yield the same results as measurement of the classical ensemble. Some students then conclude that the two ensembles are identical. While productive applications of NMP can be easily imagined, most research shows it used unproductively.

This same paper~\cite{meyer2021states} discusses another strategy students use to solve quantum computing problems, which they call `Virtual Quantum Computer.' In this strategy, students apply gates, measurements, or other features of a quantum circuit to a given state as if that state were going through a quantum computer. We observe this strategy in our data; however, when we refer to this strategy throughout our dataset, we refer to it as `playing quantum computer' to emphasize the strategy as an action students can take to solve problems.

Playing quantum computer has similarities with the concept of `plug-and-chug' expressed throughout PER. A 2016 paper on student difficulties with expectation values~\cite{singh2016expectation} found that students rarely used conceptual reasoning to interpret statements about expectation values in quantum mechanics, instead resorting to formal calculations. The authors suggest that ``even upper-level undergraduate and graduate students often prefer `plug and chug' methods as opposed to developing a coherent conceptual understanding''. While `playing quantum computer' may at first appear to be a direct analog of `plug-and-chug', it is in fact representative of the algorithmic thinking that underlies computer science and is essential to working with complicated quantum circuits. Throughout this paper, we note instances when playing quantum computer is strictly necessary to solve a problem, more efficient than conceptual reasoning, and/or less efficient than conceptual reasoning---and instances students both play quantum computer and reason conceptually at the same time.

\section{Context and Methods}\label{sec:methods}

\subsection{Interviews}\label{sec:interview-questions}

We conducted think-aloud interviews with students using questions from the Quantum Computing Conceptual Survey~\cite{meyer2025thesis, molly2025validation} (QCCS). Our sample comes from students who had taken the QCCS as part of a recent quantum computing class and who indicated that they wanted to be contacted for follow-up studies.
We scheduled 30 interviews with these students, conducted over Zoom by the first author. One of the interviews was removed from the data because they did not answer all the questions, so our total $N$ is 29 students. For transparency, we report the demographics of our sample in Table \ref{table:demographics}, but we do not disaggregate our data by demographic categories. We numbered students in order of interview date and time, and we refer to students using this number as `Student \#'.

Data collection resulted in two main artifacts per interview: detailed notes by the first author during each interview and an audio recording later transcribed via Otter.ai. The transcriptions were reviewed and corrected using the recorded video and audio files with particular attention to cases where the automated transcription failed to capture `jargon' terms. The audio recording for interview 13 failed, so any references to interview 13 are based on the first author's notes.

Interviews involved seven questions taken from the QCCS and an eighth, open-ended question written for the interview. Some of the questions were selected to further investigate their face validity in the QCCS; others were chosen to characterize student understanding of the CNOT gate. Students responded fully and without intervention to each question before the interviewer asked prompting questions, meant to further characterize the student's reasoning or suggest a different line of reasoning. (When students activate a resource as part of their response to a prompting question, we specify that they were `prompted.') After the interviews, we selected three of the four questions involving the CNOT gate for analysis;\footnote{The fourth question was broad and open-ended, but unfortunately did not yield as much insight as we hoped, so we do not include it in this analysis.} these questions are described in detail below.

\subsubsection{Question 1: Which states are unchanged by CNOT?}\label{sec:question-unchanged}

\begin{figure}[b]
\includegraphics[height=120px]{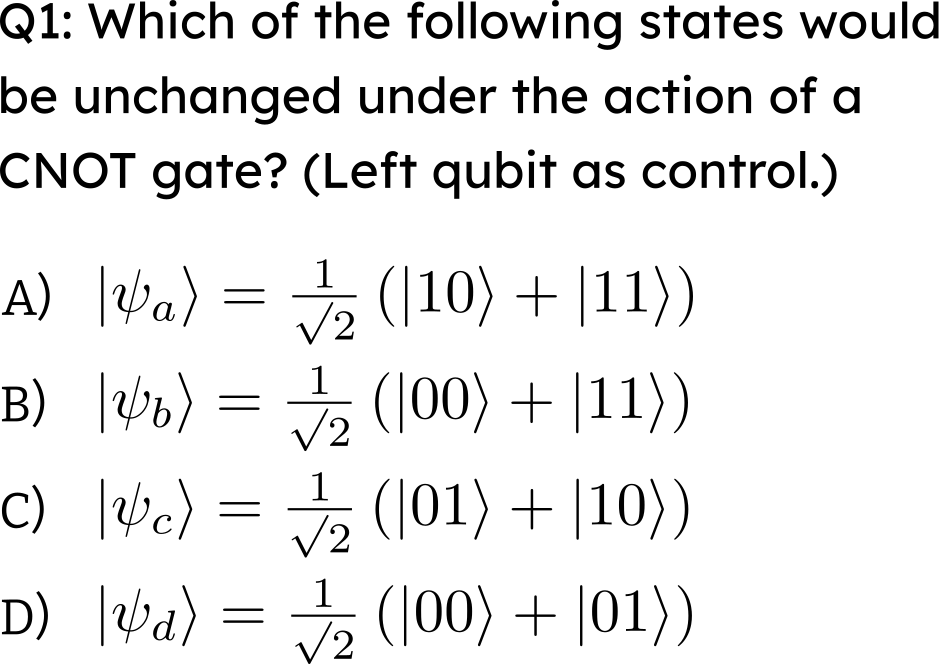}
\caption{\label{fig:question-unchanged}The prompt and answer options for Question 1. The answers are A and D, obtained by linearly applying CNOT to each individual ket.}
\end{figure}

As shown in Fig.~\ref{fig:question-unchanged}, Question 1 provides four 2-qubit quantum states and asks students which states (if any) would be left unchanged after the application of the CNOT gate. The definition of CNOT, as given in Sec.~\ref{sec:CNOT-basics}, can be applied linearly to each ket in the superposition to find that $\ket{\psi_a}$ and $\ket{\psi_d}$ are unchanged. Students must pay attention to the state as a whole, and not focus only on individual terms, because $\ket{\psi_a}$ \textit{appears} to change if one only looks at the first ket, $\ket{10}$, but even though each term changes individually, they do so in a way that keeps the state the same. Nineteen (of 29) students correctly identified which states were changed and unchanged without prompting from the interviewer.

\subsubsection{Question 2: What happens to the control qubit?}\label{sec:question-kickback}

\begin{figure}[]
\includegraphics[height=170px]{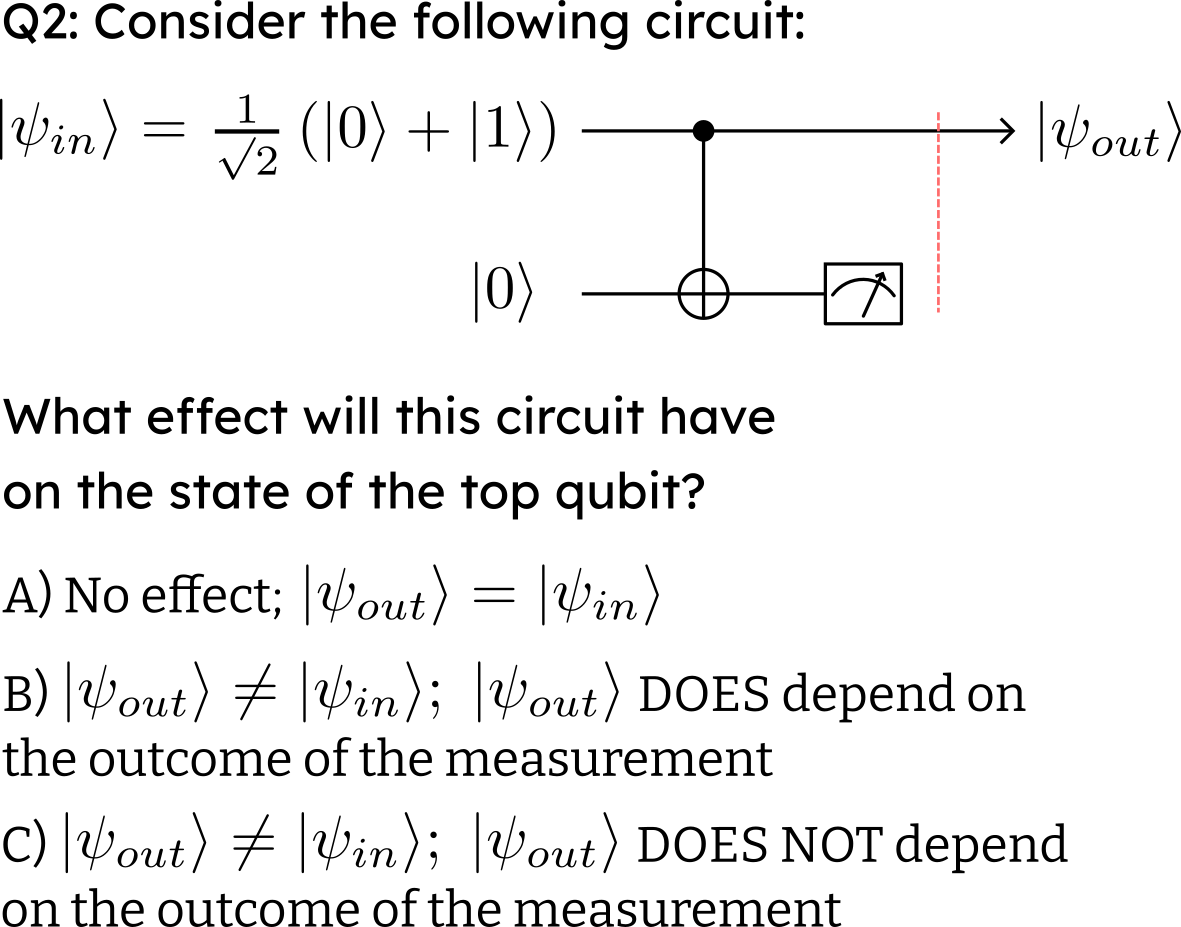}
\caption{\label{fig:question-kickback}The prompt and answer options for Question 2. The answer to this question is B because the circuit entangles the two given states, creating measurement correlation. For the given states, knowing the measurement outcome of the bottom qubit guarantees the same measurement outcome for the top qubit.}
\end{figure}

Question 2 (Fig.~\ref{fig:question-kickback}) presents students with a two-qubit circuit involving a CNOT gate and a measurement. The top qubit -- the control for the CNOT gate -- is $\ket{\psi_{in}}=\frac{1}{\sqrt{2}}\left(\ket{0}+\ket{1}\right)$. The bottom qubit, $\ket{0}$, passes through the target of that CNOT gate, and is subsequently measured in the computational basis $\{\ket{0},\ket{1}\}$. Students are asked whether the top qubit has been affected by the circuit and, if so, does the state depend on the outcome of the measurement?

A critical element of Question 2 is that the CNOT gate entangles the two qubits into the state $\ket{\psi}=\frac{1}{\sqrt{2}}\left(\ket{00}+\ket{11}\right)$. This fundamentally already represents a change to the state of the control qubit as it cannot be factored out of the two-qubit state into its original form. Measurement of the bottom ket will then collapse this two-qubit state into either $\ket{00}$ or $\ket{11}$ with $50\%$ probability. Thus, if the value measured for the target qubit is consistent with $\ket{0}$, then the control qubit must be $\ket{0}$ with $100\%$ probability. Therefore, the control qubit does change after the measurement of the target qubit and it does depend on the outcome of that measurement.

Sixteen (of 29) students answered Question 2 correctly, without prompting, though only two students
articulated without prompting that the state of the control qubit changes after the CNOT gate, not just after the measurement.

\subsubsection{Question 3: How do multiple CNOTs cancel?}

\begin{figure}[]
\includegraphics[height=141px]{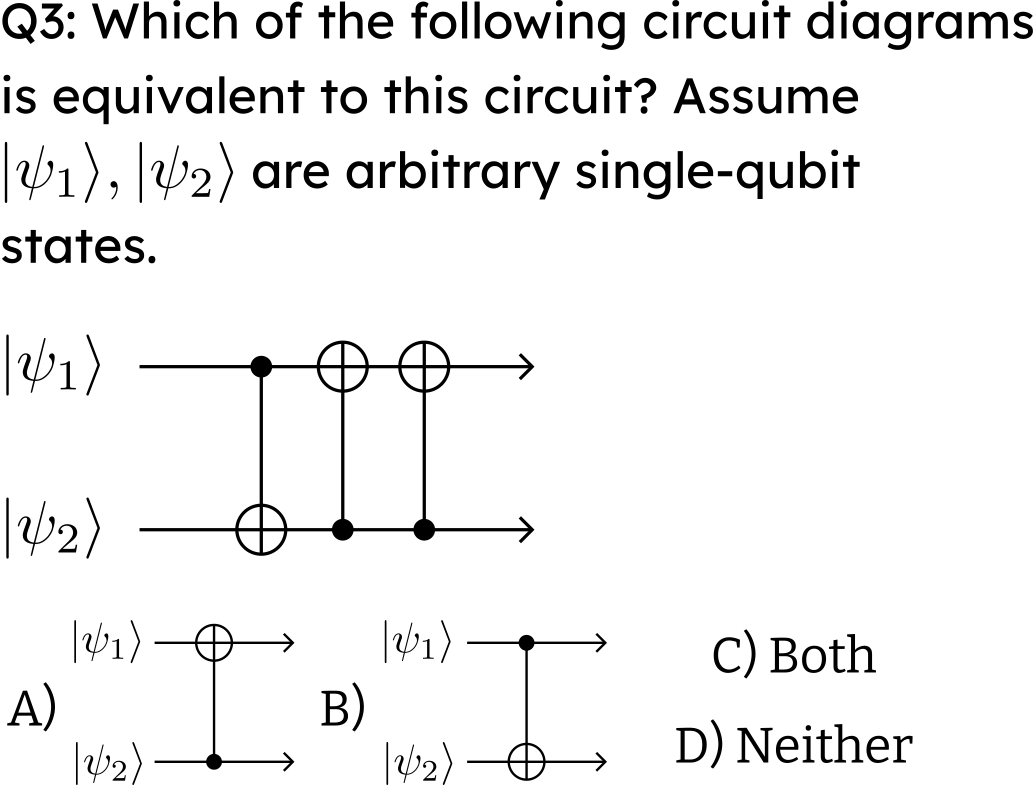}
\caption{\label{fig:question-cancellation}The prompt and answer options for Question 3. The answer to this question is B because CNOT is its own inverse; thus, the second and third CNOT gates cancel each other out.}
\end{figure}

Question 3 (see Fig.~\ref{fig:question-cancellation}) presents students with a two-qubit circuit containing three CNOT gates in various orientations. It asks students to determine whether the three-CNOT circuit is equivalent to a single-CNOT circuit (in either possible orientation), both, or neither. Two adjacent CNOT gates in the three-qubit circuit have the same orientation; the CNOT gate is its own inverse, so these two CNOTs cancel out, leaving behind one CNOT gate which can then be compared to the answer options. All except one student (out of 29) answered this question correctly without prompting.

\subsection{Theoretical Framework: Conceptual and Procedural Resources}\label{sec:theoretical-frameworks}

To explore the ways in which students reason about CNOT gates in these questions, we took a resource-based approach to framing student reasoning~\cite{hammer2000student}. Resources are often understood as individual, reusable pieces of knowledge~\cite{scherr2007modeling}. These pieces can be of many grain sizes, and may even be made up of other resources~\cite{wittmann2015mathematical}, which come together in schema-like structures~\cite{smith1994schema} that relate ideas, experiences, and contexts. Resources are \textit{activated} by students; that is, in some settings and not others, we can observe that a resource is active in a student's reasoning process by noting whether they say phrases or take actions that indicate the presence of that resource in their reasoning~\cite{barth2023methods}. When activated, resources are \textit{context-dependent}; their application to one problem may look different from their application to another problem, even while the same basic piece of knowledge is still at play~\cite{scherr2007modeling}.

We used this framework to analyze student response patterns by identifying pieces of knowledge that build up their reasoning involving the CNOT gate. All interviewees were familiar with the CNOT gate, having taken a quantum computing course in the prior semester.  Following Robertson et al.~\cite{robertson2023conceptual}, we focused on identifying resources that were `instructionally significant', i.e. related to correct physics concepts but also to the actual language students use. Instructionally significant resources can be drawn on or refined by instruction to generate physics understanding. To this end, we discarded resource candidates that were strictly incorrect in all circumstances, opting instead for resources that were, at worst, `indeterminate'~\cite{scherr2007modeling}, i.e. not necessarily correct in all contexts, but correct in some.

We consider two types of resources from the literature: conceptual resources~\cite{robertson2023conceptual} and procedural resources~\cite{wittmann2015mathematical}. To illustrate the distinction between the two types of resources, let us explore two different ways a student could answer question 2. A conceptual argument made by student 8 can be paraphrased as: `I know CNOT entangles two qubits, and I know that measuring one qubit in an entangled state causes the other qubit to collapse, so I think that in this circuit, the control qubit will collapse into either $\ket{0}$ or $\ket{1}$.' A more procedural method of solving the question might involve evaluating the circuit for the initial two-qubit state, calculating the effect of CNOT and the measurement of that state, and thereby determining that the control qubit collapses into either $\ket{0}$ or $\ket{1}$. In both cases, the student `steps through' the circuit, but the conceptual argument is more qualitative, while the procedural method is specific to the circuit's input and the calculation required to get the output.

We can now point out conceptual and procedural resources in the example arguments above. Two candidates for conceptual resources in the first example could be `CNOT entangles qubits' and `entangled qubits experience measurement correlation.' These conceptual resources are ideas that can be used to construct a qualitative or general description of the action of a circuit. In contrast, in the second example, two candidates for procedural resources could be `applying CNOT to a two-qubit state' and `determining the effect of measurement on a state'. These procedural resources are ``mathematical actions''~\cite{wittmann2015mathematical} used algorithmically, to obtain an output state from an input state.

We can distinguish these two types of resources based on their use by students and the apparent purpose they serve in a student's argument. This is true even when a procedural and conceptual resource are very closely related, such as `determining the effect of measurement on a state' and `entangled qubits experience measurement collapse'. A student can arrive at measurement correlation either by correctly determining the effect of measurement on a given entangled state or by recalling that a given state is entangled and that entanglement is characterized by measurement correlation. It is even possible that a student could use both resources at the same time.

\subsection{Analysis}

We used an emergent coding scheme to analyze student reasoning, beginning by coding common student response patterns and ending with a codebook composed of candidates for CNOT-related resources and more general QIS-related pieces of knowledge. Initially, an undergraduate researcher coded the interviews to identify modes of reasoning that appeared in multiple students' responses. The results of this coding pass suggested that we use a knowledge-in-pieces framework. The first author re-coded the data using more than a dozen pieces of knowledge that seemed to contribute to students' response patterns, and then the research team began a collaborative effort to iterate on these codes using clear exemplars and edge cases. Codes that were either too broad or too vague to be consistently applied, or too context-dependent to be useful for instruction beyond the question in which they appeared, were refined or rejected.

This process resulted in codes for three knowledge pieces related to the CNOT gate that fit our criteria for resources: clearly defined, distinct from the other resources, correct in at least some circumstances, context-dependent, and instructionally significant. It also resulted in codes for some non-CNOT-related knowledge pieces that helped facilitate our analysis but were not distinctly CNOT-related resources. These latter knowledge pieces matched those found in the broader literature---including measurement correlation, non-separability, and naive measurement probability.

These codes facilitated our analysis of student reasoning, allowing us to break up student arguments into constituent pieces that students appeared to activate. We did not perform an interrater reliability test, as we are not making claims about frequencies of resource usage. This work identifies resources and provides an exemplar-based discussion of their use by our students. We nevertheless report frequencies for transparency about whether or not we are reporting singular or very rare events; these frequencies should not be interpreted to represent generalizable results. 

\section{Results: Resources about the CNOT gate}\label{sec:defining-resources}

In this section, we describe three resources students use when working with the CNOT gate and provide examples of how they are used in context\
In the following section (Sec.~\ref{sec:discussion}), we discuss how students employ these resources and the relationships between them.

\subsection{Resource A: Applying CNOT to Specific States}\label{sec:def-resource-1}

In the context of procedural methods, where students perform calculations to extract information about a physical situation, we identify \textit{applying CNOT to specific states} as a procedural resource. The following quotations all immediately precede the use of this resource:
\begin{itemize}
    \item \emph{``So if I was to act CNOT on this one [a state in question 1]...''} (Student 7)
    \item \emph{``So the way I would think of it is, I would take any random qubit and see what would work. I'm [first] doing a $\ket{10}$ qubit.''} (Student 5)
    \item \emph{``I actually don't remember this properly. Let me just do it on the paper.''} (Student 8)
    \item \emph{``To solve this question, I think that I will probably want to do that with numbers because I wouldn't really feel too comfortable making an educated guess on this one.'' } (Student 22)
\end{itemize}
These quotations suggest that students treat this resource as a mathematical action from which they can extract information, even if they're not sure how to make an ``educated guess.'' We observed students using this resource in two ways: to determine the effect of CNOT on specific states given by the problem or to determine the effect of CNOT on `test states' that were not given by the problem (such as in question 3). These usages fit our understanding of procedural methods as `input to output' processing.

This resource (applying CNOT to specific states) is the only procedural resource we identified as directly related to the CNOT, but it is not the only procedural resource we saw evidence of in our interviews. We are considering this particular resource as one in a larger set of resources that we refer to collectively as \textit{playing quantum computer} (defined in Sec.~\ref{sec:quantum-research}). When playing quantum computer, students evaluate the effects of single gates, measurements, or entire circuits on qubit states. This evaluation may be done in a variety of ways, for example, using matrix representations or Dirac notation, which we chose not to distinguish in our analysis. Regardless of the methods students used to evaluate the output of a quantum circuit, we identified those methods as `playing quantum computer' when the narration focused on finding an output to the circuit given an input. It is possible to use conceptual resources while playing quantum computer to interpret the procedural steps one is making, and we describe instances where we observe both in Sec.~\ref{sec:res-b-qualitative}.

\subsection{Resource B: The Control-Target Definition of CNOT}\label{sec:def-resource-2}

When reasoning conceptually about the effect of the CNOT gate in circuits or on states, students often used a resource we call \textit{the control-target definition of CNOT}. In contrast to a matrix or truth table definition of CNOT,
this definition qualitatively describes CNOT's effects on basis states: `when the control qubit is $\ket{0}$, the CNOT does nothing to the target qubit, but when the control is $\ket{1}$, the CNOT flips the target'. One student who invoked the control-target definition said,
\begin{quote}
    \emph{“What [CNOT] usually does from my knowledge is, if the control is 0, it wouldn't do anything to the target. And if it was 1, it would flip the target like a NOT gate.”} (Student 4)
\end{quote}

While this resource can be used procedurally (to perform resource A), in this study, we focus on its use as a conceptual resource in students' qualitative or general arguments.\footnote{For that reason, we didn't code resource B when students used it in a procedural way---that is, to get an output state from a specific input state. Such instances were coded as resource A. But students using resource A did not always need the control-target definition, and students using the control-target definition were not always doing so procedurally, as we show in \ref{sec:res-b-qualitative}.} For example, in question 3 (Fig.~\ref{fig:question-cancellation}), many students (5 unprompted, 6 prompted, out of 29)
used this resource to qualitatively describe the effect of two CNOT gates with identical orientations and concluded that the two CNOTs `reverse' themselves. We observed students using words like `flip' as a critical part of their arguments to generalize the effect of CNOT without applying it to any specific states at all.  

\subsection{Resource C: CNOT Doesn't Change the Control Qubit}\label{sec:def-resource-B}

Implicit in the control-target definition of CNOT is CNOT's lack of action on the control qubit. When applied to pure computational basis states, CNOT may change the target qubit, but it does nothing to the control qubit. We observed students activating this implicit feature of CNOT using phrases like `\textit{CNOT doesn't change the control qubit}' (separately from the control-target definition) and in different contexts from those in which they activated the control-target definition. For this reason, and because this resource focuses solely on the control qubit while resource B primarily focuses on the target qubit or the two-qubit state as a whole, we identify it as a separate resource from resource B.

While resource C arises naturally from the definition of the CNOT, it is not universally true. When CNOT entangles or disentangles qubits, the control qubit does indeed change, so resource C cannot be used generally for all states. Nevertheless, twelve students (of 29) invoked this resource unprompted in question 2 (Fig.~\ref{fig:question-kickback}) to reason about the control qubit despite it being in a superposition state. In that question, students are asked to determine whether a qubit has changed after it has passed through the control of a CNOT gate. Regardless of what happened to the target qubit, these students reasoned that the control qubit could not have changed by invoking resource C. As explained in Sec.~\ref{sec:question-kickback}, this is not correct, and we explore the consequences of this reasoning in Sec. \ref{sec:res-c-strong}.


\section{Discussion}\label{sec:discussion}

In this section, we discuss how students solve our three interview questions using a `toolbox' composed of the three resources about the CNOT listed in the previous section, describing the contexts in which they activated these resources, whether they used these resources to solve a problem initially or check their work, and the interplay between these resources. When it is relevant, we refer to other QIS resources and strategies documented in prior literature. In our toolbox analogy, students can reach for and use various tools as part of their problem-solving process. They may choose to use some tools and not others, either through explicit choice, instinct, or a lack of comfort with certain tools. We explore patterns in students' tool use and resource activation, whether productive or unproductive, in the sections that follow.

\subsection{Students reliably apply resource A}\label{sec:res-a-productive}

All three questions in this analysis can be solved by starting with a procedural application of CNOT to specific states (resource A)---or, more generally, by playing quantum computer. The linear, step-by-step nature of playing quantum computer leaves little room for more than careless errors during the process of computation. Almost every student who used this resource did so correctly, reliably calculating the output of each state, although doing so does not guarantee a correct solution.

In two of our questions, knowledge gained from resource A led students directly to the correct answers. When answering question 1 (Fig.~\ref{fig:question-unchanged}), 19 students (of 29) fully calculated the output of all four superposition states under the action of CNOT. They could immediately compare the output states to the corresponding input states to correctly determine which states had changed and which had not. To answer question 3 (Fig.~\ref{fig:question-cancellation}), at least 5 students wrote a truth table for the three-CNOT circuit, evaluating its effect on all four computational basis states (resource A), and then they compared that truth table to the effect of a single CNOT in either orientation. This direct comparison immediately and reliably provided these students with the correct answer, but it was a lengthy process compared to other ways students answered this question (which we discuss in Sections \ref{sec:res-b-qualitative} and \ref{sec:res-a-supportive}).

In response to question 2  (Fig.~\ref{fig:question-kickback}), 24 students (13 unprompted, 11 prompted) reliably applied CNOT to the specified input state to get the entangled state $\frac{1}{\sqrt{2}}\left(\ket{00}+\ket{11}\right)$. In this instance, resource A does not guarantee success, as students must compare the output state to the input state. We discuss this further in Sec.~\ref{sec:res-c-strong}.

Procedural methods make up a crucial tool students can use to solve problems. Resource A forms the bedrock of our CNOT toolbox, because it provides trustworthy information without risking conceptual errors. However, conceptual resources can help students solve certain problems more efficiently and quickly than they could with procedural methods alone.

\subsection{Students qualitatively describe circuits using resource B}\label{sec:res-b-qualitative}

As discussed above, students can solve question 3 (Fig.~\ref{fig:question-cancellation}) by using resource A. They either apply the gates to all four computational basis states or to only one or two, and observe which answer is compatible with their calculations. But the effect of the gates in question 3 can be predicted without using resource A by observing that CNOT is its own inverse, so that the latter two identical CNOT gates cancel each other out. 

Students (11 of 29) conceptually argued that CNOT is its own inverse by using the control-target definition of CNOT (resource B). Student 20 explained the effect of the latter two CNOT gates, as follows: ``When they [the two CNOTs] are using the same control... you change something and then you change it back." Other students went into more detail, explaining how, whether the control qubit is $\ket{1}$ or $\ket{0}$, the two CNOTs perform the same reversible action (or non-action) twice. In these cases, they did not compute the effect of CNOT on specific states (and thus were not utilizing resource A), but were instead using their qualitative understanding of CNOT's behavior to analyze a more complicated scenario.

Two of these students began analyzing question 3 (Fig.~\ref{fig:question-cancellation}) using resource A and then spontaneously realized during their calculation that they could describe the behavior of the circuit using the control-target definition of CNOT (resource B). Student 8 began to apply the circuit to the four computational basis states, and while applying it to $\ket{01}$, he used the control-target definition, saying of the second and third CNOTs that ``it will change the first qubit to $\ket{1}$ and it will change the first qubit back to $\ket{0}$. So basically just reversed what it did.'' He evaluated the other states and finally repeated, ``basically, the two CNOT gates after the first one... they're reversing.'' In the middle of his procedural methods he found conceptual reasoning that he used to explain his final answer to the question. Student 22 did something similar, using the control-target definition to evaluate each computational basis state by describing how the target qubit `flips' when the control is a $\ket{1}$. After she went through all four states, she said ``going through it step by step kind of helps me see what is really happening here... it's helpful to find a pattern in it,'' and then provided a conceptual solution to the problem based on the control-target definition of CNOT (resource B), using the same `flip' wording she used to evaluate the effect of the circuit in the first place. 

This particular result---in which a conceptual argument rises spontaneously out of purely procedural reasoning---compelled us because of the `aha' moment it gave students. It is consistent with a phenomenon in the literature called `blended processing'~\cite{kuo2013students,eichenlaub2019blending}, in which students make conceptual or qualitative interpretations of their equations or calculations. In this case, students appear to blend a procedural and conceptual usage of the control-target definition: i.e., while using it as a procedural resource to apply CNOT to specific states, they understand their mathematical action in a conceptual way that they can leverage as a conceptual resource to qualitatively describe the circuit.

This phenomenon illustrates how students can build up conceptual resources from the procedural tools of quantum computing. As students applied the CNOT gate procedurally, they observed a pattern that they formed into a conceptual argument. Instruction that encourages students to reason conceptually while applying procedural tools like playing quantum computer may bring about more of these `aha' moments and build up a conceptual toolbox grounded in the bedrock of procedural resources.

\subsection{Students use resource A and other procedural tools when they are uncertain about conceptual arguments}\label{sec:res-a-supportive}

In the previous section, we observed students forming a conceptual argument after using procedural methods. This can go the other way: when a student is uncertain about their conceptual arguments, they can turn to their more reliable procedural tools to support or justify their reasoning. We present some clear exemplars below, but we do not report frequencies due to the inherent ambiguity in identifying students as `uncertain'.

When answering question 3 (Fig.~\ref{fig:question-cancellation}), some students were not certain which of their conceptual ideas was correct. Student 20 said, ``I never remember if the way that CNOTs cancel each other is when they're like in the same position or like in different positions'', and applied CNOT to the state $\frac{1}{\sqrt{2}}\left(\ket{00}+\ket{10}\right)$ to correctly determine which of these was correct. Similarly, student 29 said, ``I don't know if opposite CNOTs cancel," and after further thinking, decided he needed to ``make a logic table'', which led him to the correct answer. In these cases, resource A helped students reject incorrect statements like ``opposite CNOTs cancel''.


Other students answering question 3 (Fig.~\ref{fig:question-cancellation}) were somewhat certain that identical CNOTs canceled but were unwilling to trust that resource on faith. Student 2 said they were `pretty sure CNOT is its own inverse' and then applied CNOT to the states $\ket{01}$, $\ket{10}$, and $\ket{11}$ to conclude they could just `delete the second and third CNOTs'. Student 18 said ``I feel like I remember that'' CNOT is its own inverse, but she checked by using a different procedural tool: multiplying two CNOTs together as matrices. Student 12 started by using resource A, but when asked if he could come up with a conceptual argument for his answer after he finished his truth table, he said ``I had a feeling'' that the second and third CNOTs canceled and that that was ``lowkey my first assumption,'' but that ``my examples were just a way to make that into a concrete thing.'' Student 17 articulated an idea similar to student 12's, saying she did the math because ``it's more definitive... I don't have to confuse myself,'' even though she also demonstrated conceptual understanding when asked. 

Some students started exploring potential conceptual arguments but, unable to find one they seemed comfortable pursuing, they switched to using procedural methods instead. Consider the following examples:
\begin{itemize}
    \item Student 28, answering question 2: \emph{``So my, my gut instinct is that there's no effect, because we're not really doing anything to the top qubit. But let me think about it for a little bit. ... So why don't I just write down the action of the circuit step by step.''}
    \item Student 22, answering question 3: \emph{``To solve this question, I think that I will probably want to do that with numbers because I wouldn't really feel too comfortable making an educated guess on this one.''}
    \item Student 18, answering question 1, after some time trying to reason through the problem: \emph{``Okay, actually, I think I just have to do the math.''}
\end{itemize}

In these examples, students supported the conceptual side of their toolbox with their robust procedural tools, especially applying CNOT to specific states (resource A). Procedural methods like playing quantum computer can make students more confident about their conceptual understanding of a quantum circuit, without forfeiting or undercutting their attempts to achieve conceptual understanding. 

\subsection{Students can use resource B unproductively}\label{sec:res-b-unproductive}

While resource B was a powerful conceptual tool for students in question 3, it was very unproductive in question 1. This question (Fig.~\ref{fig:question-unchanged}) asks which of four superposition states are left unchanged by CNOT. This specific framing led 9 students to reason conceptually and generally using the control-target definition of CNOT (resource B) before looking at any individual states.\footnote{A tenth student used the matrix definition of CNOT to decide, before looking at any states, that only states with a control of $\ket{0}$ would be unchanged.} Paraphrased, the argument is as follows: `when CNOT is applied to a state with a control qubit of $\ket{1}$, it changes the target qubit, which means the state changes. Any state with a $\ket{1}$ for a control qubit must therefore be changed by CNOT.' Armed with this generalization, students looked through all four options and eliminated any which had a $\ket{1}$ for the control qubit in one or both terms in the superposition. These students answered that only the state $\frac{1}{\sqrt{2}}\left(\ket{00}+\ket{01}\right)$ was unchanged by CNOT, which, while correct, misses another unchanged option: $\ket{\psi_a}=\frac{1}{\sqrt{2}}\left(\ket{10}+\ket{11}\right)$.

Activation of resource A would have helped these students refute their incorrect generalization. Two students (of the nine who eliminated the state $\ket{\psi_a}$ using resource B) ended up checking their work unprompted using resource A,
and then they discovered their error. Another six had to be prompted to check their answers,
five (of those six) used resource A as a result, and one of those five still did not notice the error in her reasoning.
While the prompted students had access to procedural tools like resource A, they did not use them spontaneously. Checking one's conceptual reasoning about a gate or circuit in quantum computing using procedural methods is a straightforward and often quick procedure. It appears that students do not always activate procedural resources like `playing quantum computer' to check their work, but it is unclear from our data why. 


\subsection{Students trust resource C over most conflicting information}\label{sec:res-c-strong}

Question 2 (Fig.~\ref{fig:question-kickback}) requires multiple tools besides the CNOT gate in order to arrive at the correct answer. Sometimes, the conclusions students make using different tools are in conflict with each other, and students must resolve the inconsistency. In question 2, we observed 12 students using resource C (CNOT doesn't change the control qubit) to arrive, at least initially, at an incorrect conclusion: nothing happens to the control qubit in the circuit, and therefore it has not been changed. Some students checked this answer spontaneously, while others received prompting. The interviewer's prompts encouraged students to play quantum computer, whether generally (`solve the question using math') or step-by-step to indirectly elicit specific resources related to entanglement (e.g. `what is the state of the top qubit after the CNOT gate?' which was meant to elicit non-separability).

We analyzed each case individually to understand why some students changed their final answer while others did not. All 6 of the students who arrived at the correct answer activated measurement correlation at some point, and one even noticed that the state changes after the CNOT gate and before the measurement, due to non-separability. 5 of the remaining 6 students never activated measurement correlation, even when prompted. The 6th of those, who activated both non-separability and measurement correlation, ultimately rejected the correct answer because of her certainty that `CNOT doesn't change the control qubit.' We explore these arguments in more detail below.

While invoking non-separability is not necessary to solve question 3, it is nevertheless a critical part of the definition of entanglement--that is, an entangled multi-qubit state cannot be written as a tensor product of multiple single-qubit states. But the consequences of non-separability were not often well understood by the students who activated it. Students 17 and 22 both attempted to separate the entangled state $\frac{1}{\sqrt{2}}\left(\ket{00}+\ket{11}\right)$ and realized they could not, but both those students eventually concluded that the top qubit must still be in the state $\frac{1}{\sqrt{2}}\left(\ket{0}+\ket{1}\right)$. Student 17 first said,
\begin{quote}
    \emph{``We have to describe [the single qubit state] as part of the larger quantum system. ... We can't just say what the state is like, you can't just say that it's in... an equal superposition of $\ket{0}$ and plus $\ket{1}$.''}
\end{quote}
But then, re-invoking her incorrect conclusion based on resource C, she decided,
\begin{quote}
    \emph{``$\psi_{out}$ would still be an equal superposition... because I'm pretty confident about this answer, because I do still believe it [the circuit] would have no effects,''}
\end{quote}
For student 7, non-separability was a source of confusion. When asked what the state of the top qubit was after the CNOT gate, they said,
\begin{quote}
    ``You can't really treat this as a one-particle problem to begin with. ...  would say, yes, it’s the same state, if I were to, like, close my eyes and pretend the second particle didn’t exist, but it’s not really the same state because they’re entangled. Despite the fact that the control gate does, like doesn’t do anything to this top particle, entangling it kind of in effect does do something.''
\end{quote}
Ultimately, they changed their incorrect conclusion because of measurement correlation, but they could not settle on a clear interpretation of their inability to factorize the entangled state in this question.

These observations suggest that some students may be able to recognize that `something is up' when they cannot factorize a multi-qubit state, but may not know how this fits in with their understanding of single-qubit states. This corroborates findings that students struggle with the practical difference between pure and mixed states as found in Refs. \cite{passante2015states} and \cite{meyer2021states}. Alternatively, the presence of this difficulty could mediate between the contradictory findings by Zwickl and Hersom~\cite{zwickl2024entanglement} and by Kohnle and Deffebach~\cite{kohnle2015entanglement} regarding how students relate non-separability to entanglement, but more research is required to understand how students interpret non-separable states.

It is not surprising that a resource like `CNOT doesn't change the control qubit' could override a student's reasoning based on a more difficult concept like non-separability, but resource C was also strong enough to bypass a concept as fundamental to QIS as measurement correlation. Student 13 trailed off while determining the result of the measurement and invoked resource C, and he had to carefully play quantum computer to discover his error. Student 17 affirmed that after measurement, the system would ``collapse'' into either $\ket{00}$ or $\ket{11}$, but still felt ``pretty confident about this [incorrect] answer'', arguing that ``$\ket{\psi_{out}}$ would still be an equal superposition''. The phrasing of ``equal superposition'' suggests that she may have been using naive measurement probability~\cite{meyer2021states}, since the probability of measuring $\ket{00}$ or $\ket{11}$ is $50\%$, but she did not explicitly say this. Student 12 used it more obviously--when asked to write down the ``combined two-qubit state'' after the CNOT gate, he did so correctly and invoked NMP to say that the top qubit is in the state $\frac{1}{\sqrt{2}}\left(\ket{0}+\ket{1}\right)$ because ``there would still be a fifty-fifty chance''. Both these students had concluded that, by the end of the circuit, the system was either in a $\ket{00}$ or $\ket{11}$ state, but they resolved the contradiction with resource C by maintaining that the qubit was still in a quantum superposition.

In contrast, two students (5 and 22) did not invoke measurement correlation whatsoever. Student 22 observed that the state was ``a Bell state'', but continued,
\begin{quote}
    \emph{``However, even though it's a Bell state $\ket{\psi}$ still represents the leftmost qubit in this entire state. And even if you perform a measurement on the rightmost qubit, you're still going to have your leftmost qubits be $\frac{1}{\sqrt{2}}\left(\ket{0}+\ket{1}\right)$''}
\end{quote}
Student 22 understood collapse in Dirac notation for a single qubit, but did not demonstrate understanding of how it applied to a multi-qubit state, especially one she identified as a Bell state. It appears that she did not conceptually activate measurement correlation despite mentioning `Bell state'---it is possible that she does not associate `Bell states' with entanglement, or entanglement with measurement correlation. 

Nevertheless, five of the seven students
who explicitly invoked measurement correlation used it to correct their answer. Student 28 said, ``my gut instinct is that there's no effect, because we're not really doing anything to the top qubit'', but then he spontaneously played quantum computer until he arrived at the measurement with the correct entangled state in hand. ``In computational basis,'' he began, trailing off and presumably evaluating the measurement. He then concluded at once that his ``gut instinct'' was wrong and that the answer ``will depend on the measurement''. The success of these five students may represent a fluency in using Dirac notation to predict the effects of measurement on superpositions and multi-qubit states.

We did not anticipate how strongly students would stick to resource C even when they had access to other basic, fundamental quantum computing concepts like measurement correlation, but it seems to be a natural generalization for students to make without realizing the contexts in which it fails. It is the sort of generalization a student could easily make after applying CNOT to two-qubit kets many times in their QIS education and never changing the control qubit. In order for students to build up a robust quantum computing toolbox composed of both procedural tools and conceptual generalizations, they must be encouraged to check their generalizations procedurally and to determine the range of applicability of their conceptual ideas.

\section{Conclusion and Future Research}\label{sec:conclusion}

\subsection{The CNOT Toolbox: Implications for Quantum Computing Instruction}

In the above discussion, we have characterized students' CNOT gate `toolbox' as it appears in our data. While students were perfectly capable of using the CNOT gate procedurally to answer questions, they also used the procedural application of the CNOT gate to justify and strengthen their conceptual ideas about CNOT. They generalized the effect of CNOT into two conceptual resources that they applied throughout our questions, both productively and unproductively, and sometimes, but not always, checked their conceptual answers by playing quantum computer. 

Our observations suggest that procedural tools (like playing quantum computer) may form the basis of students' quantum computing toolbox. All three questions could be solved purely by using procedural tools; indeed, in questions 1 and 2, procedural tools were almost always more effective than conceptual resources, which students tended to overgeneralize. But in question 3, students' use of a conceptual resource was much more efficient than procedural methods and revealed a general pattern that students using purely procedural methods were less likely to notice.

Though this study presented some issues with students' conceptual generalizations, we nevertheless believe that students would benefit greatly from a practice of developing and testing conceptual tools so that they can understand quantum circuits and algorithms on a more abstract or qualitative level. The `testing' component is crucial here---instructors should encourage students to check their ideas using procedural methods to understand in which contexts their conceptual ideas apply. In quantum computing, the fluent use of both procedural resources and conceptual resources, as needed or in tandem, represents expert-like behavior. It is not simply `plug-and-chug' to play quantum computer. It is foundational and necessary, and it can be illuminating when leveraged to develop strong and useful conceptual ideas that facilitate qualitative analysis of circuits and algorithms in quantum computing.

\subsection{Limitations and Avenues for Future Research}

Selection effects apply when interpreting the results of our study. All students in our study had taken an introductory quantum computing course previously, and it has been shown that these courses are primarily housed at large, private R1 institutions~\cite{meyer2024disparities}. As such, and because we are not making any statistical generalizations from our study, our results should not be seen to characterize the entirety of student reasoning regarding CNOT. 

This study was designed to elicit CNOT-related reasoning in a variety of circumstances in order to build up an understanding of the tools students use when working with the CNOT gate, but it was not designed to target the specific resources we found. Now that we have found three CNOT-related resources, we encourage future researchers to design studies aimed at those specific resources, whether to understand how students form these resources, to observe students using them in various contexts, or to target the problems that arise from misuse of these resources. Some other avenues for future research follow, based on non-CNOT-related observations from our dataset. 

We observed that the strategy of `playing quantum computer' (called `virtual quantum computer' by Meyer et al.~\cite{meyer2021states}) is reliable for evaluating circuits in quantum computing. It is a foundational strategy that represents the algorithmic thinking of computer scientists. Knowing when to play quantum computer to test a theory or gain information is an expertlike behavior and often makes the difference between an incorrect conceptual answer and a correct, fully justified answer. Future avenues for research could include explorations of: Do students and faculty see playing quantum computer as expertlike behavior? Do they conceptually interpret their calculations (i.e., engage in blended processing~\cite{kuo2013students,eichenlaub2019blending})? 

To play quantum computer effectively, students also need facility with Dirac notation, especially as it relates to multi-qubit states. While research has discussed whether students can recognize entanglement from states in Dirac notation when prompted to look for it, there is little to no research on how often students spontaneously activate resources related to entanglement when working with arbitrary states. Future research could explore students' fluency in recognizing entangled states from Dirac notation, reading measurement correlation from Dirac notation, and whether Dirac notation helps students interpret the non-separability of entangled qubits (some work has been done already on student interpretations of non-separability~\cite{brang2024spooky}).

\section{Acknowledgments}

This work was eased by an REU student, Michael Burnes, who helped correct the transcriptions and did an initial pass categorizing common student response patterns. The questions used in this work were supplied by the Quantum Computing Conceptual Survey (QCCS), developed mainly by Josephine Meyer. She also provided insight into how our results fit into the landscape of QIS education research. This work was supported by NSF Grants No. 2143976, 2011958, and 2012147.




\bibliography{CNOTResources}

@article{meyer2024introductory,
  title={Introductory quantum information science coursework at US institutions: content coverage},
  author={Meyer, Josephine C and Passante, Gina and Pollock, Steven J and Wilcox, Bethany R},
  journal={EPJ Quantum Technology},
  volume={11},
  number={1},
  pages={16},
  year={2024},
  publisher={Springer Berlin Heidelberg}
}

@inproceedings{plunkett2020survey,
  title={A survey of educational efforts to accelerate a growing quantum workforce},
  author={Plunkett, Thomas and Frantz, Terrill L and Khatri, Hamida and Rajendran, Praveen and Midha, Sunny},
  booktitle={2020 IEEE International Conference on Quantum Computing and Engineering (QCE)},
  pages={330--336},
  year={2020},
  organization={IEEE}
}

@inproceedings{cervantes2021overview,
  title={An overview of quantum information science courses at {US} institutions},
  author={Cervantes, Bianca and Passante, Gina and Wilcox, Bethany R and Pollock, Steven J},
  booktitle={Physics Education Research Conference Proceedings},
  year={2021}
}

@article{meyer2022today,
  title={Today’s interdisciplinary quantum information classroom: Themes from a survey of quantum information science instructors},
  author={Meyer, Josephine C and Passante, Gina and Pollock, Steven J and Wilcox, Bethany R},
  journal={Physical Review Physics Education Research},
  volume={18},
  number={1},
  pages={010150},
  year={2022},
  publisher={APS}
}

@article{asfaw2022building,
  title={Building a quantum engineering undergraduate program},
  author={Asfaw, Abraham and Blais, Alexandre and Brown, Kenneth R and Candelaria, Jonathan and Cantwell, Christopher and Carr, Lincoln D and Combes, Joshua and Debroy, Dripto M and Donohue, John M and Economou, Sophia E and others},
  journal={IEEE Transactions on Education},
  volume={65},
  number={2},
  pages={220--242},
  year={2022},
  publisher={IEEE}
}

@article{dzurak2022development,
  title={Development of an undergraduate quantum engineering degree},
  author={Dzurak, Andrew S and Epps, Julien and Laucht, Arne and Malaney, Robert and Morello, Andrea and Nurdin, Hendra I and Pla, Jarryd J and Saraiva, Andre and Yang, Chih Hwan},
  journal={IEEE Transactions on Quantum Engineering},
  volume={3},
  pages={1--10},
  year={2022},
  publisher={IEEE}
}

@article{marshman2015framework,
  title={Framework for understanding the patterns of student difficulties in quantum mechanics},
  author={Marshman, Emily and Singh, Chandralekha},
  journal={Physical Review Special Topics-Physics Education Research},
  volume={11},
  number={2},
  pages={020119},
  year={2015},
  publisher={APS}
}

@article{krijtenburg2017insights,
  title={Insights into teaching quantum mechanics in secondary and lower undergraduate education},
  author={Krijtenburg-Lewerissa, Kim and Pol, Hendrik Jan and Brinkman, Alexander and van Joolingen, Wouter R},
  journal={Physical review physics education research},
  volume={13},
  number={1},
  pages={010109},
  year={2017},
  publisher={APS}
}

@article{bouchee2022towards,
  title={Towards a better understanding of conceptual difficulties in introductory quantum physics courses},
  author={Bouch{\'e}e, Tim and de Putter-Smits, Lesley and Thurlings, M and Pepin, B},
  journal={Studies in Science Education},
  volume={58},
  number={2},
  pages={183--202},
  year={2022},
  publisher={Taylor \& Francis}
}

@article{corsiglia2023intuition,
  title={Intuition in quantum mechanics: Student perspectives and expectations},
  author={Corsiglia, Giaco and Pollock, Steven and Passante, Gina},
  journal={Physical Review Physics Education Research},
  volume={19},
  number={1},
  pages={010109},
  year={2023},
  publisher={APS}
}

@article{emigh2020based,
  title={Research-based quantum instruction: Paradigms and Tutorials},
  author={Emigh, Paul J and Gire, Elizabeth and Manogue, Corinne A and Passante, Gina and Shaffer, Peter S},
  journal={Physical Review Physics Education Research},
  volume={16},
  number={2},
  pages={020156},
  year={2020},
  publisher={APS}
}

@article{passante2024interactive,
  title={Interactive homework to support student learning of measurement uncertainty in quantum mechanics},
  author={Passante, Gina and Kohnle, Antje},
  journal={Physical Review Physics Education Research},
  volume={20},
  number={2},
  pages={020131},
  year={2024},
  publisher={APS}
}

@article{sadaghiani2015quantum,
  title={Quantum mechanics concept assessment: Development and validation study},
  author={Sadaghiani, Homeyra R and Pollock, Steven J},
  journal={Physical Review Special Topics-Physics Education Research},
  volume={11},
  number={1},
  pages={010110},
  year={2015},
  publisher={APS}
}

@article{gire2015dirac,
  title={Structural features of algebraic quantum notations},
  author={Gire, Elizabeth and Price, Edward},
  journal={Physical Review Special Topics—Physics Education Research},
  volume={11},
  number={2},
  pages={020109},
  year={2015},
  publisher={APS}
}

@inproceedings{kushimo2023dirac,
  title={Investigating students' strengths and difficulties in quantum computing},
  author={Kushimo, Tunde and Thacker, Beth},
  booktitle={2023 IEEE International Conference on Quantum Computing and Engineering (QCE)},
  volume={3},
  pages={33--39},
  year={2023},
  organization={IEEE}
}

@inproceedings{kohnle2015entanglement, Author = "Antje Kohnle and Erica Deffebach", Title = {Investigating student understanding of quantum entanglement}, booktitle={Physics Education Research Conference Proceedings},
  year={2021} }

@inproceedings{zwickl2024entanglement,
  title={Investigating Students’ Understanding of Entanglement},
  author={Hersom, Hope J and Zwickl, Benjamin M},
booktitle={Physics Education Research Conference Proceedings},
  year={2024}
}

@inproceedings{meyer2021states,
  title={Investigating students’ strategies for interpreting quantum states in an upper-division quantum computing course},
  author={Meyer, Josephine and Passante, Gina and Pollock, Steven J and Vignal, Michael and Wilcox, Bethany R},
  booktitle={Physics Education Research Conference Proceedings},
  year={2021}
}

@article{passante2015states,
  title={Student ability to distinguish between superposition states and mixed states in quantum mechanics},
  author={Passante, Gina and Emigh, Paul J and Shaffer, Peter S},
  journal={Physical Review Special Topics—Physics Education Research},
  volume={11},
  number={2},
  pages={020135},
  year={2015},
  publisher={APS}
}

@inproceedings{singh2016expectation, Author = "Chandralekha Singh and Emily Marshman", Title = {Student difficulties with determining expectation values in quantum mechanics}, BookTitle = {Physics Education Research Conference Proceedings}, Year = {2016} }

@unpublished{molly2025validation,
Author="Griston, Molly and Meyer, Josephine and Wilcox, Bethany", Title="TBD", Note="In press"}

@phdthesis{meyer2025thesis,
    author = "Meyer, Josephine",
    title = "New Paradigms in Quantum Education for the Second Quantum Revolution",
    school = "University of Colorado - Boulder",
    year = 2025
}

@phdthesis{corsiglia2023thesis,
    author = "Corsiglia, Giaco",
    title = "Designing Online Activities for Teaching Quantum Mechanics: A Research-Based Approach",
    school = "University of Colorado - Boulder",
    year = 2023
}

@inproceedings{corsiglia2022basis, Author = "Giaco Corsiglia and Steven Pollock and Bethany Wilcox", Title = {Effectiveness of an online homework tutorial about changing basis in quantum mechanics}, BookTitle = {Physics Education Research Conference}, Year = {2022} }

@article{robertson2023conceptual,
  title={Identifying student conceptual resources for understanding physics: A practical guide for researchers},
  author={Robertson, Amy D and Goodhew, Lisa M and Bauman, Lauren C and Hansen, Brynna and Alesandrini, Anne T},
  journal={Physical Review Physics Education Research},
  volume={19},
  number={2},
  pages={020138},
  year={2023},
  publisher={APS}
}

@article{smith1994schema,
  title={Misconceptions reconceived: A constructivist analysis of knowledge in transition},
  author={Smith III, John P and DiSessa, Andrea A and Roschelle, Jeremy},
  journal={The journal of the learning sciences},
  volume={3},
  number={2},
  pages={115--163},
  year={1994},
  publisher={Taylor \& Francis}
}

@article{hammer2000student,
  title={Student resources for learning introductory physics},
  author={Hammer, David},
  journal={American journal of physics},
  volume={68},
  number={S1},
  pages={S52--S59},
  year={2000},
  publisher={American Association of Physics Teachers}
}

@article{wittmann2015mathematical,
  title={Mathematical actions as procedural resources: An example from the separation of variables},
  author={Wittmann, Michael C and Black, Katrina E},
  journal={Physical Review Special Topics—Physics Education Research},
  volume={11},
  number={2},
  pages={020114},
  year={2015},
  publisher={APS}
}

@article{barth2023methods,
  title={Methods of research design and analysis for identifying knowledge resources},
  author={Barth-Cohen, Lauren A and Swanson, Hillary and Arnell, Jared},
  journal={Physical Review Physics Education Research},
  volume={19},
  number={2},
  pages={020119},
  year={2023},
  publisher={APS}
}

@article{scherr2007modeling,
  title={Modeling student thinking: An example from special relativity},
  author={Scherr, Rachel E},
  journal={American Journal of Physics},
  volume={75},
  number={3},
  pages={272--280},
  year={2007},
  publisher={American Association of Physics Teachers}
}

@article{kuo2013students,
  title={How students blend conceptual and formal mathematical reasoning in solving physics problems},
  author={Kuo, Eric and Hull, Michael M and Gupta, Ayush and Elby, Andrew},
  journal={Science Education},
  volume={97},
  number={1},
  pages={32--57},
  year={2013},
  publisher={Wiley Online Library}
}

@incollection{eichenlaub2019blending,
  title={Blending physical knowledge with mathematical form in physics problem solving},
  author={Eichenlaub, Mark and Redish, Edward F},
  booktitle={Mathematics in physics education},
  pages={127--151},
  year={2019},
  publisher={Springer}
}

@article{brang2024spooky,
  title={Spooky action at a distance? A two-phase study into learners’ views of quantum entanglement},
  author={Brang, Michael and Franke, Helena and Greinert, Franziska and Ubben, Malte S and Hennig, Fabian and Bitzenbauer, Philipp},
  journal={EPJ Quantum Technology},
  volume={11},
  number={1},
  pages={33},
  year={2024},
  publisher={Springer Berlin Heidelberg}
}

@article{meyer2024disparities,
  title={Disparities in access to {US} quantum information education},
  author={Meyer, Josephine C and Passante, Gina and Wilcox, Bethany},
  journal={Physical Review Physics Education Research},
  volume={20},
  number={1},
  pages={010131},
  year={2024},
  publisher={APS}
}

@article{robinett2002QMVI, Author = {Richard Robinett and E Cataloglu}, Title = {Testing the development of student conceptual and visualization understanding in quantum mechanics through the undergraduate career}, Journal = {Am. J. Phys.}, Volume = {70}, Number = {3}, Pages = {238-251}, Month = {March}, Year = {2002} }

\end{document}